\documentclass[twocolumn,superscriptaddress,amsfont,amssymb,amsmath, showpacs,balancelastpage, nofootinbib]{revtex4-1}
\usepackage{graphicx,longtable,mathrsfs,color,array}
\usepackage[hidelinks]{hyperref}
\usepackage[usenames,dvipsnames]{xcolor}
\usepackage{amssymb,amsmath,mathtools,mathrsfs}
\usepackage{epsfig,subfigure,placeins,float}
\usepackage{booktabs,longtable,ctable,multirow}
\usepackage{exscale,relsize}
\usepackage[normalem]{ulem}
\usepackage{enumerate}
\usepackage{times, mathptmx}
\usepackage{colortbl} 
\usepackage[dvipsnames]{xcolor} 
\usepackage{pifont} 
\usepackage{amsmath}

\newcommand{\be}{\begin{equation}}
\newcommand{\ee}{\end{equation}}

\def\MM{M_{*}}

\newcommand{\R}{R}

\newcommand{\Gtwo}{G_2{}}
\newcommand{\Gthree}{G_3{}}
\newcommand{\Gfour}{G_4{}}
\newcommand{\Gfive}{G_5{}}

\newcommand{\Hfour}{F_4{}}
\newcommand{\Hfive}{F_5{}}

\def\d{\delta}

\allowdisplaybreaks

\begin{document}
 
\title{{Dark Energy after GW170817}}
\author{Paolo Creminelli}
\affiliation{Abdus Salam International Centre for Theoretical Physics,
Strada Costiera 11, 34151, Trieste, Italy}
\author{Filippo Vernizzi}
\affiliation{Institut de physique th\' eorique, Universit\'e  Paris Saclay 
CEA, CNRS, 91191 Gif-sur-Yvette, France}
        \date{\today}

\begin{abstract}
The observation of GW170817 and its electromagnetic counterpart implies that gravitational waves travel at the speed of light, with deviations smaller than a few $\times 10^{-15}$. We discuss the consequences of this experimental result for models of dark energy and modified gravity characterized by a single scalar degree of freedom. To avoid tuning, the speed of gravitational waves must be unaffected not only for our particular cosmological solution, but also for nearby solutions obtained by slightly changing the matter abundance. For this to happen the coefficients of various operators must satisfy precise relations that we discuss both in the language of the Effective Field Theory of Dark Energy and in the covariant one, for Horndeski, beyond Horndeski and degenerate higher-order theories. The simplification is dramatic: of the three functions describing quartic and quintic beyond Horndeski theories, only one remains and reduces to a standard conformal coupling to the Ricci scalar for Horndeski theories. We show that the deduced relations among operators do not introduce further tuning of the models, since they are stable under quantum corrections.
\end{abstract}

\maketitle

\vskip.1cm
\emph{Introduction.}  The association of GW170817 \cite{TheLIGOScientific:2017qsa} and GRB 170817A \cite{Goldstein:2017mmi}  events  allowed  to make an extraordinarily precise measurement of the speed of gravitational waves (GWs): it is compatible with the speed of light with deviations smaller than a few  $\times 10^{-15}$ \cite{Monitor:2017mdv}.  This measurement dramatically improves our understanding of dark energy/modified gravity. These scenarios are characterised by a cosmological ``medium" which interacts gravitationally with the rest of matter. This medium, at variance with a simple cosmological constant, spontaneously breaks Lorentz invariance, so that there is no a priori reason to expect that gravitational waves, which are an excitation of this medium, travel at the same speed as photons \cite{Lombriser:2015sxa,Bettoni:2016mij}. 

The measurement is of particular relevance since it probes the speed of GWs over cosmological distances. The change of speed might be locally reduced in high density environments, but it is difficult to believe that this screening effect can persist over distances of order 40 Mpc.
Moreover one has to stress that this is a low energy measurement, at a scale  as low as 10 000~km. For such a low energy, one should be allowed to use the Effective Field Theory (EFT) of Dark Energy or Modified Gravity which applies to cosmological scales. Actually, in the theories we are going to study, the cutoff may be of the same order as the measured GW momentum and high-dimension operators may play some role; however one does not expect that high-energy corrections conspire to completely cancel the modification of the GW speed. On the other hand, previous stringent limits from gravitational Cherenkov radiation of cosmic rays \cite{Moore:2001bv}  are only applicable to high energy GWs, well outside the regime of validity of the EFTs describing Dark Energy and Modified Gravity. Moreover these bounds only apply to GWs travelling faster, and not slower, than light. For other  limits see \cite{Baskaran:2008za,Jimenez:2015bwa,Brax:2015dma,Blas:2016qmn}.

With these caveats in mind, in this paper we want to explore what are the consequences of this measurement in the context of the Effective Field Theory (EFT) of Dark Energy \cite{Creminelli:2008wc,Gubitosi:2012hu,Bloomfield:2012ff} and in its covariant counterpart, the Horndeski \cite{Horndeski:1974wa,Deffayet:2011gz} and the beyond Horndeski theories \cite{Gleyzes:2014dya} (see also \cite{Zumalacarregui:2013pma}). If we impose that the absence of an effect is robust under tiny variations of the cosmological history---say a small variation of the dark matter abundance---we find that one needs precise relations among the various coefficients of the operators. This allows us to derive the most general scalar-tensor theory compatible with GWs travelling at the speed of light.
Since the required relations must be satisfied with great accuracy, given the experimental precision, one needs to understand whether they are radiatively stable. We will see that they  are stable under quantum corrections due to the non-renormalization properties of these theories.

\vskip.1cm
\emph{Consequences for the EFT of Dark Energy.}  The EFT of Dark Energy is a convenient way to parametrize cosmological perturbations around a FRW solution with a preferred slicing induced by a time-dependent background scalar field.  For the time being we assume that matter is minimally coupled to the gravitational metric; we will come back to this point later on. 

Expanded around a  FRW background, $ds^2 = -dt^2 + a^2(t) d\vec x^2$ and written in a gauge where the time coincides with uniform field hypersurfaces, the EFT action reads
\be
\begin{split}
\label{total_action}
S  = & \int  d^4 x \sqrt{-g}  \bigg[ \frac{\MM^2}{2} f \, {}^{(4)}\!R - \Lambda- c g^{00}  +  \frac{m_2^4}{2} (\delta g^{00})^2 \\
&- \frac{m_3^3}{2} \, \delta K \delta g^{00}  
- m_4^2 \delta {\cal K}_2 +\frac{\tilde m_4^2}{2} \,\delta g^{00} \R   - \frac{m_5^2}{2}  \delta g^{00} \delta {\cal K}_2  \\   &-   \frac{m_6}{3} \delta {\cal K}_3  
 - \tilde m_6  \delta g^{00} \delta {\cal G}_2  -   \frac{m_7}{3} \delta g^{00} \delta {\cal K}_3  \bigg] \;.
\end{split}
\ee
Here ${}^{(4)}\!R$ is the 4d Ricci scalar, $\delta g^{00} = 1+g^{00}$, $\d K_\mu^\nu\equiv K_\mu^\nu- H \delta_\mu^\nu$  is the perturbation of the extrinsic curvature of the time hypersurfaces ($H\equiv \dot a/a$), $R_\mu^\nu$ is the 3d Ricci tensor of these hypersurfaces, and $\delta K$ and $R$ are respectively their trace.
For convenience we have also defined
\be
\begin{split}
\delta {\cal K}_2 &\equiv \delta K^2 -  \d K_\mu^\nu \d K^\mu_\nu \;, \qquad \delta {\cal G}_2  \equiv  \delta K_\mu^\nu R^\mu_\nu -  \d K R/2 \;, \\
\delta {\cal K}_3  &\equiv \delta K^3 -3 \d K \d K_\mu^\nu \d K^\mu_\nu + 2 \d K_\mu^\nu \d K^\mu_\rho \d K_\nu^\rho   \;.
\end{split}
\ee
While $\MM^2$ is constant, the other parameters are time-dependent functions.
As we will discuss in the following section, this action describes the cosmological perturbations in Horndeski (for $\tilde m_4^2=m_4^2$ and $\tilde m_6=m_6$) and beyond Horndeski theories.  
At quadratic order,  it has been introduced in \cite{Gleyzes:2013ooa}. At higher order, we have written only the  operators  that contribute to the leading number of spatial derivatives. These dominate  the nonlinear regime of structure formation and  the Vainshtein regime (see e.g.~\cite{Kimura:2011dc,Kobayashi:2014ida} and \cite{Cusinetal2017} for details). At quintic or higher order there are no such operators. The other operators present in Horndeski and beyond Horndeski theories are not explicitly written  but will be discussed below. More general higher-order operators will be considered below.

In eq.~\eqref{total_action}, GWs only enter in the 4d and 3d Ricci tensor and in the trace-free part of $K_\mu^\nu$. At quadratic order, the operator $m_4^2 \delta {\cal K}_2$ contributes to the graviton kinetic energy, changing the normalization of the effective Planck mass---which becomes $M^2 \equiv \MM^2 f +2 m_4^2$---modifying the propagation speed of gravitational waves \cite{Cheung:2007st,Gleyzes:2013ooa},
\be
c_{T}^2 -1 =  - {2 m_4^2}/{M^2}  \;.
\ee
(Notice that $m_4^2$ can have either signs, it is written as a square just to keep track of dimensions.)
Thus, the constraint of GW170817 implies that the coefficient of the operator $m_4^2 \delta {\cal K}_2$ must be extremely small,
\be
\label{cond1}
m_4^2=0\;.
\ee

However, the value of this parameter depends on the particular background the EFT is expanded around. In particular, by changing by a tiny amount the Hubble expansion or  the background energy density of the scalar (or, correspondingly,  the dark matter abundance) the coefficients of the EFT action get reshuffled. A change in the background appears in the EFT action as a background value for $\delta g^{00}$ and $\delta K$.
To robustly set to zero $m_4^2$ we should set to zero also all those operators that can generate it by a small change of the background solution. As an example, consider  $  m_5^2 \delta g^{00} \d {\cal K}_2$. When $\delta g^{00}$ is evaluated on the background, this operator becomes quadratic and shifts the parameter $m_4^2$,  i.e., $\delta m_4^2 = m_5^2 \delta g^{00}_{\rm bkgd}/2$. However, the change in $c_{T}^2$ can be compensated by the operator $ \tilde m_4^2 \delta g^{00} R$ if $\tilde m_4^2$ is chosen appropriately. By choosing 
\be
\label{cond2}
\tilde m_4^2 = m_5^2  \qquad ( = 0 \;{\rm in\;Horndeski)}\;,
\ee
these two operators combine to change the overall normalization of the graviton action, keeping the graviton on the light-cone. (In Horndeski: $m_4 = \tilde m_4=0$.)
The same tuning must hold for operators with more powers of $\delta g^{00}$ that have not been explicitly included in the action, such as $(\delta g^{00})^2 \R $, $(\delta g^{00})^2 \delta {\cal K}_2 $, etc.

Let us consider the remaining operators, starting with $m_6 \delta {\cal K}_3$. When one of the $\delta K_\mu^\nu$ or $\delta K$ in the cubic expression for $\delta {\cal K}_3$ is evaluated on the background, this operator becomes quadratic and contributes to $ m_4^2$. Using $(\delta K_\mu^\nu)_{\rm bkgd} = \delta H_{\rm bkgd} \delta_\mu^\nu $ one finds  $\delta m_4^2 = \delta H_{\rm bkgd} m_6$. Notice that the dependence on the background is through $\delta H_{\rm bkgd}$ and not through  $\delta g^{00}_{\rm bkgd}$, so that its contribution cannot be compensated by  neither $\tilde m_4^2$ nor $m_5^2$. It is easy to get convinced that the same happens for $\tilde m_6$ and $m_7$. When $\delta g^{00}$ is evaluated on the background, upon use of eq.~(8) of \cite{Gleyzes:2013ooa} one finds that the operator $\tilde m_6$ shifts $m_4^2$ by $\delta m_4^2 = - \frac12 (\tilde m_6 \delta g^{00}_{\rm bkgd})^{\hbox{$\cdot$}} $. Finally, the operator $m_7$ induces $\delta m_4^2 =  m_7 \delta g^{00}_{\rm bkgd} \delta H_{\rm bkgd} $. Since the background enters differently in all these operators, they must be precisely set to zero,
\be
\label{cond3}
m_6 = \tilde m_6 = m_7 =0 \;.
\ee
As we will discuss below, the relations we found are stable under radiative corrections. 

\vskip.1cm
\emph{Covariant action.} 
Let us see how the constraints of GW170817 on the EFT of Dark Energy translate for covariant theories. In particular, we consider the action 
\be
S= \int d^4 x \sqrt{-g} \sum_I L_I\;,
\ee
where we have defined the Lagrangians
\be
\begin{split}
L_2 & \equiv  \Gtwo(\phi,X)\;,   \qquad L_3  \equiv  \Gthree(\phi, X) \, \Box \phi \;,  \\
L_4 & \equiv \Gfour(\phi,X) \, {}^{(4)}\!R - 2 \Gfour_{,X}(\phi,X) (\Box \phi^2 - \phi^{ \mu \nu} \phi_{ \mu \nu}) \\
& -\Hfour(\phi,X) \epsilon^{\mu\nu\rho}_{\ \ \ \ \sigma}\, \epsilon^{\mu'\nu'\rho'\sigma}\phi_{\mu}\phi_{\mu'}\phi_{\nu\nu'}\phi_{\rho\rho'}\;, \\
L_5 & \equiv \Gfive(\phi,X) \, {}^{(4)}\!G_{\mu \nu} \phi^{\mu \nu} \\ 
& +  \frac13  \Gfive_{,X} (\phi,X) (\Box \phi^3 - 3 \, \Box \phi \, \phi_{\mu \nu}\phi^{\mu \nu} + 2 \, \phi_{\mu \nu}  \phi^{\mu \sigma} \phi^{\nu}_{\  \sigma})  \\ \;&
 \quad- \Hfive (\phi,X) \epsilon^{\mu\nu\rho\sigma}\epsilon^{\mu'\nu'\rho'\sigma'}\phi_{\mu}\phi_{\mu'}\phi_{\nu \nu'}\phi_{\rho\rho'}\phi_{\sigma\sigma'} \label{LHBH}\,,
\end{split}
\ee
that depend on a scalar field $\phi$, $X\equiv g^{\mu \nu} \partial_\mu \phi \partial_\nu \phi$ and second derivatives of the field. For convenience, we denote the scalar field derivatives by $\phi_\mu\equiv \nabla_\mu \phi$, $  \phi_{\mu\nu}\equiv \nabla_{\nu}\nabla_\mu \phi$ and $\Box \phi \equiv \phi^\mu_\mu$. The symbol  $\epsilon_{\mu \nu \rho \sigma }$ is the totally antisymmetric Levi-Civita tensor and a comma denotes a partial derivative with respect to the argument. Horndeski theories are recovered by the conditions $\Hfour(\phi,X) = 0$ and $ \Hfive (\phi,X) =0$,
which guarantee that the equations of motion are purely second order. 
If $L_5=0$ and $G_4 - 2X G_{4,X} \neq 0$ ($L_4=0$ and $G_{5,X} \neq 0$), it is possible to go beyond Horndeski by switching on $F_4 \neq 0$ ($F_5 \neq 0$) without propagating more than one single scalar and the graviton \cite{Gleyzes:2014dya} (see also \cite{Deffayet:2015qwa,Langlois:2015cwa}). 
If both $L_4 $ and $L_5$ are present, the condition for the beyond Horndeski theories to be degenerate \cite{Langlois:2015cwa} and propagate a single degree of freedom  is 
\be
\label{deg}
X G_{5,X} F_4 = 3 F_5 \left[ G_4 - 2 X G_{4,X} - (X/2) G_{5,\phi} \right] \;,
\ee
which can be obtained by imposing that both Lagrangians are generated by the same disformal transformation \cite{Gleyzes:2014qga}. In summary, the quartic and quintic Lagrangians of beyond Horndeski theories are described in terms of three independent functions of $\phi$ and $X$

To compare with the EFT approach, let us write the relevant  parameters in eq.~\eqref{total_action}  in terms of the covariant functions $G_4$, $G_5$, $F_4$ and $F_5$ above (of course $L_2$ and $L_3$ do not affect GWs),
\be
\begin{split}
M^2 & =  2 G_4 - 4 X G_{4,X} - X \big( G_{5,\phi} + 2 H \dot \phi  G_{5,X}  \big) \\&
+ 2 X^2 F_4 -6H \dot \phi X^2 F_5\;, \\
m_4^2 & = \tilde m_4^2 + X^2 F_4 - 3 H \dot \phi X^2 F_5   \;, \\
\tilde m_4^2 & = - \big[ 2 X G_{4,X} + X G_{5,\phi} + \big(H \dot \phi - \ddot \phi \big) X G_{5,X}  \big] \;, \\
m_5^2 & = X \big[ 2 G_{4,X} +   4 X G_{4,XX} + H \dot \phi  ( 3 G_{5,X}  + 2  X  G_{5,XX})  + G_{5,\phi}  \\ & + X G_{5,X \phi}   - 4 X F_4 -2X^2 F_{4,X}  + H \dot \phi  X \big( 15 F_5 + 6 X F_{5,X} \big) \big] \;,\\
m_6  & = \tilde m_6 -   3 \dot \phi X^2 F_{5}      \;,\qquad  \tilde m_6  = -  \dot \phi X  G_{5,X}  \;, \\
 m_7 & = \frac{1}{2} \dot \phi X \big(3 G_{5,X}+2 X G_{5,XX} +15 X F_{5} + 6 X^2 F_{5,X} \big) \;. \label{alphasGF}\end{split}
\ee

Setting the speed of GWs to one, i.e., eq.~\eqref{cond1}, implies that the particular combination appearing in the expression of $m_4^2$ above vanishes. This must be true on any background  and thus must hold for any value of  $\ddot \phi$,  $H$ and $\dot \phi$ (or $X$). This  implies, respectively,
\be
\label{back_ind}
G_{5,X}=0\;, \qquad F_5=0 \;, \qquad  2 G_{4,X}  - X F_4 + G_{5,\phi} =0 \;,
\ee
for any $X$ and $\phi$.
Thus, $G_5$ can be at most a function of $\phi$, the beyond Horndeski term $F_5$ must be absent and there is a relation between $G_{4,X}$ and $F_4$ and their derivatives. 
The first two conditions automatically imply eq.~\eqref{cond3}. It is also straightforward to verify that eq.~\eqref{cond2} is a consequence of eq.~\eqref{back_ind}.
Finally, using eq.~\eqref{back_ind} in $L_4$ and $L_5$ of the Lagrangians \eqref{LHBH}, after some manipulations and integrations by parts  we remain with
\be
\label{eq:BHredux}
\begin{split}
& L_{c_T=1}=     G_2(\phi,X) + G_3(\phi,X) \Box\phi+ B_4(\phi,X) \, {}^{(4)}\! R \\
&- \frac4X B_{4,X} (\phi,X) (\phi^\mu \phi^\nu \phi_{\mu \nu} \Box \phi - \phi^\mu \phi_{\mu \nu} \phi_\lambda \phi^{\lambda \nu}) \;,
\end{split}
\ee
where we have defined $B_4 \equiv G_4 + X G_{5,\phi}/2 $.
To show that this theory does not change the speed of tensors we can decompose the 4d Ricci using the Gauss-Codazzi relation and after some integration by parts one finds
\be
L_{c_T=1} =  G_2 + G_3 \Box\phi+B_4 ( R +  K_\mu^\nu K^\mu_\nu - K^2  ) \;,
\ee
where $K_\mu^\nu$, $K$ and $R$ are respectively the extrinsic curvature tensor, its trace and the 3d Ricci scalar of the uniform $\phi$ hypersurfaces.
Note that from eq.~\eqref{back_ind} $ 2 B_{4,X} = X F_4$. Thus, in the absence of a beyond Horndeski operator, $F_4=0$, the second term in this equation vanishes and $B_4$ is only a function of $\phi$ so that we recover a standard conformal coupling to the 4d Ricci scalar, i.e.,  $ B_4(\phi) \, {}^{(4)}\! R$. 

So far, we have assumed that $c_T=1$ is robust under independent variations of $H$, $\dot\phi$ and $\ddot\phi$: indeed both the expansion history and $\phi(t)$ change if one modifies, for instance, the dark matter abundance. This however does not happen in the particular cases when dark energy has a fixed $\dot\phi$ independently of $H$. In the EFT language one can check that the change in $g^{00}$ induced by a change $\delta H_{\rm bkgd}$ is of order $c/(c + 2 m_2^4) \cdot \delta H_{\rm bkgd}/H$. If $c=0$ (and therefore $\Lambda$ in eq.~\eqref{total_action} is time-independent) the variation of the cosmological history does not give rise to a change in $\dot\phi$. Notice that dark energy acts like a cosmological constant at background level. 
In this case, the condition $m_4^2=0$ does not automatically require that $G_{5,X}$ and $F_5$ vanish independently but it only requires that they are related by $G_{5,X} + 3 X F_5=0$, and only on the attractor solution. However, this condition together with the degeneracy equation \eqref{deg} and $m_4^2=0$ imply the pathological value $M=0$, unless $G_{5,X}$ and $F_5$ separately vanish. In the EFT language one still has $m_6 = \tilde m_6 =0$, but in general $\tilde m_4^2 \neq m_5^2$ and also $m_7$ is independent.

\vskip.1cm
\emph{Radiative stability.}  We saw that the observation of GW170817 imposes, both in the EFT description and in the covariant one, some precise relations among the coefficients of various operators. Of course it is crucial to understand whether these relations are stable under quantum corrections, otherwise one would have to rely, order by order in perturbation theory, on a $10^{-15}$  tuning. Let us discuss this issue in the covariant theory. As discussed in \cite{Pirtskhalava:2015nla}, the Horndeski theories inherit some of the  properties of the Galileons \cite{Luty:2003vm},
for which the leading operators cannot be generated by loop graphs.
This strongly constraints the size of quantum corrections in our case. 

Let us assume the functions $G_4$ and $G_5$ do not depend on $\phi$ and are of the form 
\be
G_4(X) = \frac{\Lambda_2^8}{\Lambda_3^6} \hat G_4\left(\frac{X}{\Lambda_2^4}\right) \;, \qquad G_5(X) = \frac{\Lambda_2^8}{\Lambda_3^9} \hat G_5\left(\frac{X}{\Lambda_2^4}\right) \;.
\ee 
To have sizeable dark energy effects one takes $\Lambda_2 \sim (M_{\rm Pl} H_0)^{1/2}$ and $\Lambda_3 \sim (M_{\rm Pl} H_0^2)^{1/3}$, 
where $M_{\rm Pl}$ is the Planck mass. We take the dimensionless functions $\hat G$ to be polynomials in their variable with order one coefficients $c_n$. The result of \cite{Pirtskhalava:2015nla} is that  all these coefficients are corrected by a relative amount of order $\delta c_n \sim (\Lambda_3/\Lambda_2)^4 \sim 10^{-40}$. This is much smaller than the $10^{-15}$  cancellation implied by the measurement of the speed of GWs: it is completely negligible unless one goes to extraordinary large $n$. The same conclusions can be obtained in a beyond Horndeski theory \cite{SantoniTrincherini}. In conclusions the relation one has to invoke to be compatible with GW170817 are technically natural in the sense that once imposed at tree level they are stable under quantum corrections.

\vskip.1cm
\emph{Higher-Order Operators and Conformal Transformations.} It was recently pointed out that there are more general theories than those in eq.~\eqref{LHBH} that do not propagate additional degrees of freedom
\cite{Langlois:2015cwa}. In the EFT language they give rise to particular combinations of the quadratic operators \cite{Langlois:2017mxy}
\be
\int  d^4 x \sqrt{-g}  \frac{M^2}{2} \bigg(-\frac23 \alpha_L \delta K^2 +4 \beta_1 \delta K V + \beta_2 V^2 + {\beta_3} a_i a^i
\bigg)\;,
\ee
where $V \equiv -\frac12 (\dot g^{00} - N^i \partial_i g^{00})/g^{00}$ and $a_i = -\frac12 \partial_i g^{00}/g^{00}$. It is straightforward to see that these operators do not affect the speed of GWs. This is true around the given background, but also if one considers different backgrounds: since these operators have two derivatives, only $\delta g^{00}$ can be turned on, but it is easy to see that even around the new background GWs are unaffected. 

In the covariant language these theories can be obtained starting from beyond Horndeski and performing a conformal transformation that depends on $X$. Since this does not change the  light-cone, if one starts from the action (\ref{eq:BHredux}) also the resulting degenerate higher-order theories will not affect GWs speed of propagation. Under a general conformal transformation $g_{\mu \nu} \to C(\phi,X) g_{\mu \nu}$ \cite{Crisostomi:2016czh,Achour:2016rkg} we find (we assume $C$ is not linear in $X$)
\be
\begin{split}
& L_{c_T=1}=  \tilde G_2 + \tilde G_3 \Box\phi +  C B_4 \, {}^{(4)}\! R -\frac{ 4 C B_{4,X} }{X} \phi^\mu \phi^\nu \phi_{\mu \nu} \Box \phi  \\
&+ \bigg( \frac{4C B_{4,X} }{X} +\frac{6 B_4 C_{,X}{}^2}{C}+  8C_{,X} B_{4,X} \bigg) \phi^\mu \phi_{\mu \nu} \phi_\lambda \phi^{\lambda \nu} \\
&-\frac{8 C_{,X} B_{4,X}}{X} (\phi_\mu \phi^{\mu \nu } \phi_\nu)^2 \;. \label{DHOST}
\end{split}
\ee
(We do not explicitly show the expression of $\tilde G_2$ and $\tilde G_3$, since they are anyway free functions unrelated to the other terms.) This is the most general degenerate theory which can be obtained from Horndeski by a metric redefinition compatible with $c_T^2=1$. In the classification of Ref.~\cite{Langlois:2015cwa} it belongs to type Ia DHOST  theories.

There are theories in which spacial (but not time) higher derivatives are present and therefore do not propagate extra degrees of freedom. In the case of the Ghost Condensate \cite{ArkaniHamed:2003uy}, the modification of the GW speed goes as $c_T^2 - 1\sim M_{\rm gc}^2/M_{\rm Pl}^2$, where $M_{\rm gc}$ is the typical scale of the model. Since experimental bounds on the modification of the Newton law give $M_{\rm gc} \lesssim$~10 MeV, one does not expect any significant effect on the speed of GWs. On the other hand, in the case of Einstein-Aether \cite{Jacobson:2008aj} and Ho\v rava gravity \cite{Blas:2010hb}  $c_T$ is expected to deviate from unity and the bound of GW170817 represents a severe constraint on these models.

\vskip.1cm
\emph{Disformal transformations.} So far, we have assumed that matter is minimally coupled to the metric. There is no lack of generality in this, provided there is a universal coupling for all matter species, since one can always go to this frame with a suitable conformal and disformal transformation. In this frame the results of GW170817 imply that GWs must travel on the lightcone of the metric. If one chooses to go to a different disformal frame, both matter and GWs will acquire a common disformal coupling: since they both travel at the same speed, this is obviously still compatible with what LIGO/Virgo observed. 
In the new frame, the gravitational action will not be of the form \eqref{eq:BHredux} or \eqref{DHOST}.
For example, one can decide to disform the beyond Horndeski theories \eqref{eq:BHredux} to become a Horndeski theory, but now both GWs and light will not move on the geodesics of the metric.

\vskip.1cm
\emph{Conclusion.} We have obtained the most general scalar-tensor theories propagating a single scalar degree of freedom compatible with the observation of GW170817. In Jordan frame, the parameters of the EFT of Dark Energy of these theories must satisfy eqs.~\eqref{cond1}, \eqref{cond2} and \eqref{cond3}. Analogous relations must be imposed on the operators containing higher order terms in $\delta g^{00}$. The most general covariant theory is given by eq.~\eqref{DHOST}. 

After GW170817, quartic and quintic Horndeski theories are excluded, unless they reduce to a standard conformal coupling to ${}^{(4)}\!R$. Consequently, the cubic and quartic operators of eq.~\eqref{total_action} must be absent, which implies that the Vainshtein mechanism allowed by them \cite{Kimura:2011dc} cannot take place (screening must rely only on the cubic  theories) and that no signatures of these nonlinear operators should be found in the large scale structures (see e.g.~\cite{Takushima:2013foa}).
For beyond Horndeski theories, the Vainshtein mechanism is broken inside compact bodies \cite{Kobayashi:2014ida}. We leave for the future to study what consequence this has on the theories \eqref{DHOST}.

The relations that need to be satisfied are technically natural, but it would be nice to investigate whether they can be derived from some underlying symmetry. On the experimental side further observations over a larger distance and at lower frequencies will make the limits even more robust to Vainshtein screening and higher derivative corrections.

\vskip.1cm
\emph{Acknowledgements:} This paper follows up from very interesting discussions during the workshop DARK MOD, where this work was initiated. We kindly acknowledge the workshop participants and the Paris-Saclay funding. Moreover, we thank M.~Lewandowski for useful discussions and D.~Langlois and E.~Babichev 
for pointing out sign typos respectively in eq.~(16) and (8).  F.V. acknowledges financial support from ``Programme National de Cosmologie and Galaxies" (PNCG) of CNRS/INSU, France and  the French Agence Nationale de la Recherche under Grant ANR-12-BS05-0002. 
\appendix


 \bibliographystyle{utphys}
\bibliography{EFT_DE_biblio3}

\end{document}